\documentclass[a4paper,11pt]{article}
\pdfoutput=1 

\usepackage{jinstpub} 

\title{Level Zero Trigger Processor for the NA62 experiment}

\author[a,1]{Dario Soldi,\note{Corresponding author}}
\author[b]{Stefano Chiozzi}

\affiliation[a]{Istituto Nazionale di Fisica Nucleare - Universita' di Torino, via Pietro Giuria 1, 10137 Torino, Italy}
\affiliation[b]{Istituto Nazionale di Fisica Nucleare - Polo Scientifico Tecnologico - Universita' di Ferrara - via Saragat 1 - 44122 - Ferrara, Italy}

\emailAdd{dario.soldi@to.infn.it}

\abstract{The NA62 experiment is designed to measure the ultra-rare decay $K^+ \rightarrow \pi^+ \nu \bar{\nu}$ branching ratio with a precision of $\sim 10\%$ at the CERN Super Proton Synchrotron (SPS). 
The trigger system of NA62 consists in three different levels designed to select events of physics interest in a high beam rate environment.  
The L0 Trigger Processor (L0TP) is the lowest level system of the trigger chain. It is hardware implemented using programmable logic. The architecture of the NA62 L0TP system is a new approach compared to existing systems used in high-energy physics experiments. It is fully digital, based on a standard gigabit Ethernet communication between detectors and the L0TP Board. The L0TP Board is a commercial development board, mounting a programmable logic device (FPGA). The primitives generated by sub-detectors are sent asynchronously using the UDP protocol to the L0TP during the entire beam spill period. The L0TP realigns in time the primitives coming from seven different sources and performs a data selection based on the characteristics of the event such as energy, multiplicity and topology of hits in the sub-detectors.
It guarantees a maximum latency of 1 ms. The maximum input rate is about 10 MHz for each sub-detector, while the design maximum output trigger rate is 1 MHz. A description of the trigger algorithm is presented here.
}


\begin{document}
\maketitle
\flushbottom

\section{Introduction} 
NA62 is an experiment designed to study rare kaon decays exploiting the proton beam at the CERN Super-Proton Synchrotron (SPS) accelerator\cite{Anelli}\cite{DetectorPaper}. The NA62 main goal is to measure the ultra-rare decay $K^+ \rightarrow \pi^+ \nu \bar{\nu}$ branching ratio with a precision of $\sim 10\%$. 
Given the theoretical precision with which this ratio is calculated\cite{pinunu}, this decay is one of the best probes for new physics complementing the direct searches performed within the Non-Minimal Flavor Violation theory.

NA62 uses a decay-in-flight  technique to measure kaon decays, matching well with the characteristics of the CERN SPS facility.
The incoming beam derives from a primary 400 GeV/$c$ proton beam impinging on a beryllium target. Positive particles with a 75 GeV/$c$ momentum are selected by a system made of magnets and collimators. The high beam intensity and the energy of the particles do not allow to separate kaons from the other products coming from the target.
Thus the kaon tagging is performed with a Cherenkov detector characterized by a time resolution of $\sim$70 ps.
Then NA62 is equipped with two spectrometers, one placed in the detector upstream region, providing high precision time, direction and momentum of all incoming beam particles, and the other one placed after the decay region to provide position and momentum measurement of the decay products.
The upstream spectrometer, called GigaTracker, consists of three stations of hybrid silicon pixel detectors installed  around four dipole magnets.    
The downstream spectrometer consists of four straw chambers and a dipole magnet located between the second and third chamber.
The magnetic field imparts a horizontal transverse momentum kick of 270 MeV/$c$ to charged particles.
Immediately downstream of the third station of the GigaTracker, a veto counter detecting the inelastic interactions of the beam particles in the last station material has been installed.
Muon and pion tracks with momenta in the range from 15 to 35 GeV/$c$ are separated using a Ring Imaging Cherenkov (RICH) detector.
Two hodoscopes (NA48-CHOD and CHOD) are useful to determine time and position of the charged tracks.
A hermetic photon veto system ensures a coverage at angles up to 50 mrad for photons coming from the decay region. 
It has been designed to suppress the background due to $K^+ \rightarrow \pi^+\pi^0 (\pi^0 \rightarrow \gamma\gamma)$ and radiative decays by identifying photons with an inefficiency less than $10^{-8}$.
The Large Angle Veto (LAV) system covers the region between 8.5 and 50 mrad. 
The Liquid Krypton calorimeter (LKr), inherited from the NA48 experiment and equipped with a new readout, covers an acceptance from 1.0 to 8.5 mrad and should guarantee an inefficiency lower than $10^{-5}$
for photons with energy less than 35 GeV.
Small-angle calorimeters (IRC, SAC) complement the detection at angles  from 0 to 1.0 mrad.
Finally, a muon veto system consisting of three different components (MUV1, MUV2 and
MUV3) is placed after the LKr calorimeter. 
Two hadronic iron/scintillator sampling calorimeters  (MUV1,2) are used to measure the deposited energies and shower shapes of incident particles and redundantly distinguish muons from pions.
A scintillator-tile muon detector (MUV3) is placed behind MUV2 and a 80 cm thick iron wall. It can be used also in the particle identification.

\begin{figure}[h] 
\centering
\includegraphics[width=0.6\textwidth]{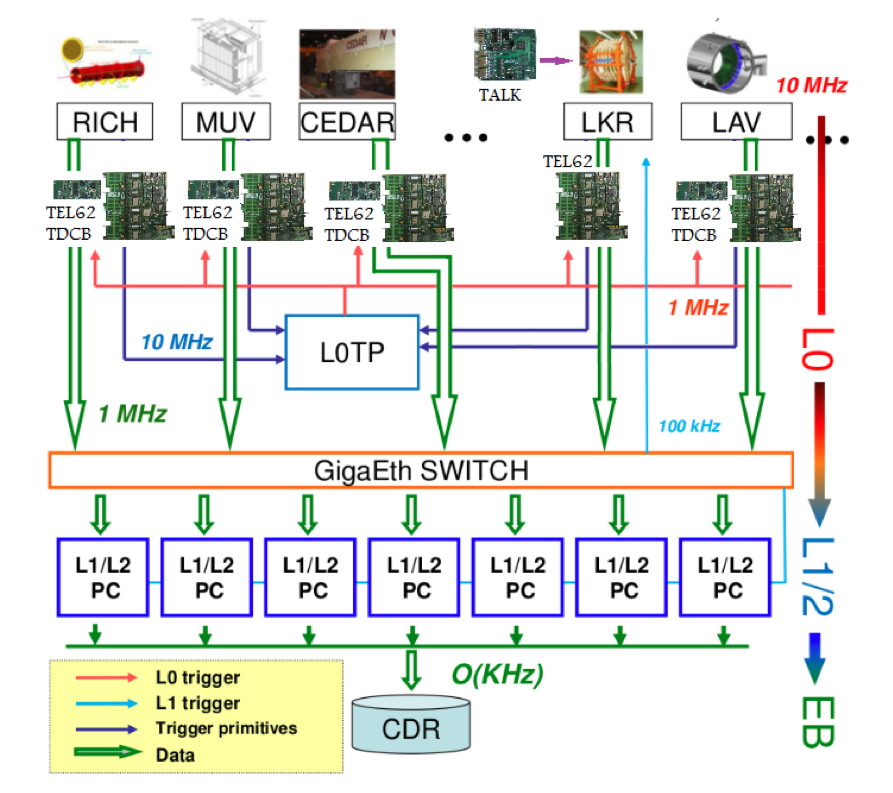}
\caption{The schematic represents an overview of the NA62 trigger system. The information, stored in primitives, is generated in a subset of detectors and transmitted to L0TP, where is processed in order to reduce the rate from 10 MHz to 1 MHz. Data satisfying the L0 selection are moved from all the detectors to the PC-Farm, where high level triggers (L1 and L2) are applied, reducing the rate from 1 MHz to about 20 kHz. Only the events passing the three selections, L0, L1, and L2, are written on disk by the Central Data Recording (CDR) service.}
\label{fig:picScheme}
\end{figure}

\paragraph{Trigger and Data Acquisition} 
Searches of ultra-rare processes require a high-instantaneous beam intensity to collect sizable signal event samples.
For this reason, one of the main challenge for high intensity experiments such NA62 is the design of the trigger and data acquisition system. 
The event rate requires a trigger system able to guarantee a high acceptance for the signal events, while keeping a high rejection of the background.
Together with the main trigger selecting $K^+ \rightarrow \pi^+ \nu \bar{\nu}$ events, other special triggers have been implemented to collect data in order to explore exotic kaon decays. 
\\
The NA62 trigger hierarchy is made of three logical levels. 
The first level is a hardware L0 trigger, based on the input from a subset of detectors. After a positive response from L0, data are readout from front-end electronics buffers, packed into Multi-Event-Packets (MEP) and sent to the PC-Farm, where a software L1 trigger, based on information computed independently by each sub-detector system is performed. A software L2 trigger, based on assembled and (partially) reconstructed events, defines the final sample, which is stored on disk by the Central Data Recording (CDR) service for further analysis.  
The scheme of the system is shown in figure \ref{fig:picScheme}.

\section{L0 Trigger Processor}
\begin{figure}[!h] 
\centering
\includegraphics[width=.6\textwidth]{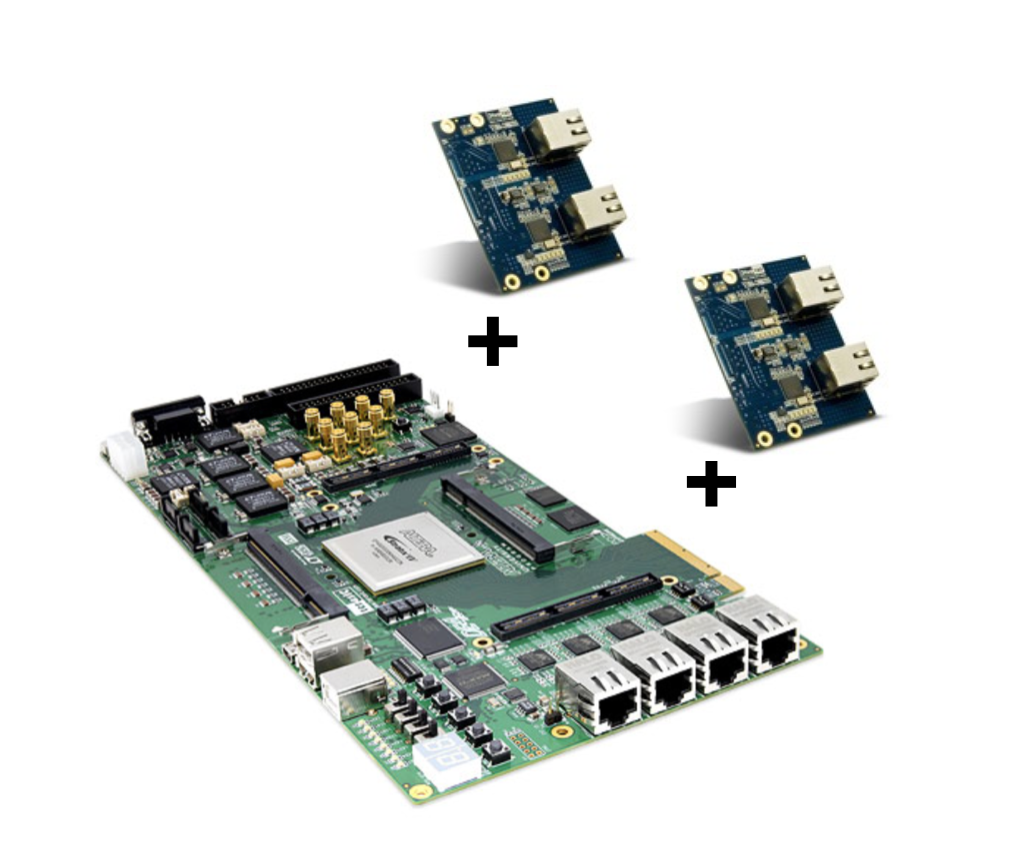}
\caption{The picture shows the Terasic DE4 board and the two HSMC-NET Daughter Boards.}
\label{fig:picBoard2}
\end{figure}

The SPS accelerator delivers the beam in spills of five seconds, called \textit{bursts}, with a maximum of three bursts per minute. 
During the burst, each of sub-detectors concurring to form the L0 triggers produces local information, encoding in the so-called \textit{primitive} the characteristics of the event and the time of occurrence.
Primitives are sent to the L0TP that has to process data coming from a maximum of seven different sources.
Up to 10 MHz of primitives per detector can be handled by the system. 
\\
Whenever the trigger conditions are fulfilled, the L0TP sends a signal synchronously with the NA62 clock to the all detectors, in order to record the sub-detector information.
Moreover, it sends an Ethernet packet containing all the characteristics of the trigger to the NA62 PC-Farm, in order to store the trigger conditions offline.
In NA62, the L0 output bandwidth is saturated at 1 MHz. 
\\
The L0TP algorithm is implemented on a commercial development board, Terasic DE4\cite{DE4},  shown in figure \ref{fig:picBoard2}, mounting an Altera Stratix IV FPGA\footnote{Stratix IV GX EP4SGX530KH40C2}\cite{Altera}. 

\begin{figure}[h] 
\centering
\includegraphics[width=0.6\textwidth]{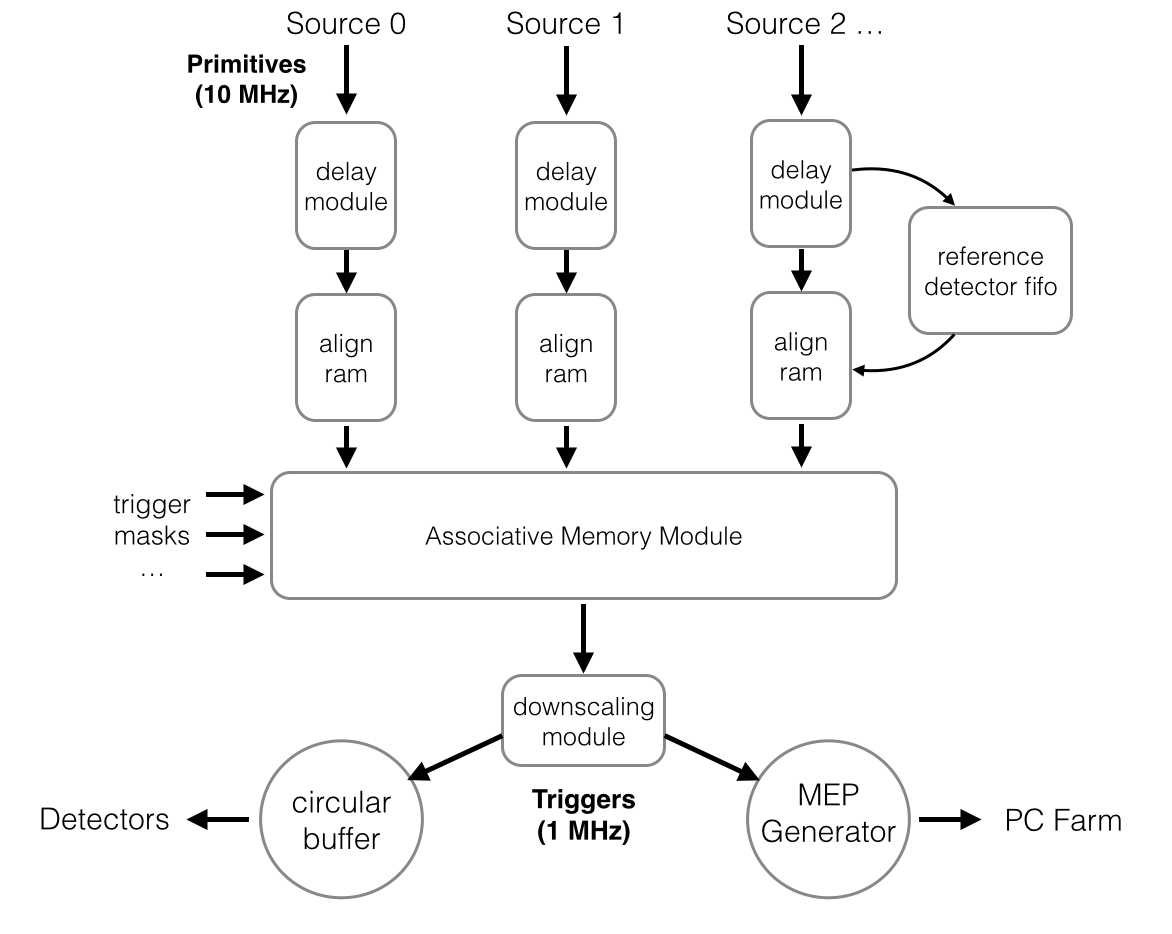}
\caption{The sketch shows a simplified scheme of the L0 Trigger Processor firmware. The L0TP receives primitives from different sources, with an average rate of 10 MHz. As a first step, the  Delay Module compensates the eventual offset between different ethernet packets in which the primitives are stored. The L0TP realigns then the primitives in time using the timing information to generate the address of the memory buffers, one for each source. During the reading process of the buffers, a specific detector acts as a reference source: the memory addresses of the reference detector primitives are stored into a dedicated FIFO, marking the only memory locations which will be read for all sources. All the other sub-systems are checked to be in time with respect to the reference detector in a certain time window, and then their conditions will be compared with the pre-selected trigger masks implemented in the Associative Memory Module. If a trigger masks is satisfied, the output trigger is generated and sent, eventually after being scaled down, to all the NA62 detectors, while the trigger information is dispatched to the NA62 PC-Farm.}
\label{fig:picTriggerScheme}
\end{figure}

The DE4 development board hosts four on-board Ethernet nodes connected with serial SGMII interfaces to the FPGA. Four additional Ethernet nodes are added hosting two Terasic HSMC-NET Daughter Boards, which are connected to the FPGA with a RGMII parallel interface.
With these extensions, the DE4 board allows to have 8-gigabit Ethernet links. 
In the NA62 implementation, seven links have been dedicated to the sub-detectors and the remaining one to output the results.
The DE4 board provides also two DDR2 SO-DIMM sockets, giving the possibility to have up to 8 GB of external memory clocked at 400 MHz.
However, because the Stratix IV FPGA is characterized by 820K logic elements (LEs), 23.1 Mb of embedded memory, and up to 1,288 18 x 18 multipliers, the entire logic of the trigger has been implemented without using any external memory.
\\
An auxiliary board has been implemented to interface the L0TP with the NA62 master clock, detectors and SPS control signals. 
The auxiliary board hosts a TTCex board\cite{TTC}, which receives the clock signal via optical  fiber and translates it to LVDS standard, accepted by the DE4. Similarly, it receives the start signal and the end signal of the beam-extraction delivered by the SPS.
On the other side, the auxiliary board receives the output from the L0TP and dispatches it to all the detectors, which are connected in daisy-chain.  
Finally, the auxiliary board receives choke/error signals coming from all the NA62 detectors, thus the data taking can be suspended when a choke/error condition has been detected.
\\
A simplified scheme of the firmware logic is shown in figure \ref{fig:picTriggerScheme}.

The logic in the L0TP is driven by two clock domains: a 40 MHz clock, received from the common NA62 time distribution system, and a 125 MHz clock, driving the trigger-algorithm logic and the Ethernet communication.
The 40 MHz clock drives the entire data acquisition system and is the common reference for all time measurements in NA62. As a consequence, all the L0TP outputs are synchronized with this clock.
The 125 MHz clock comes from a PLL fed with the 50 MHz DE4 internal oscillator input. 
Special care was required to handle the two clock domains, using dual-clock FIFO buffers and multiple synchronization flip-flops.
A common reset for the different clock domains is also provided to clear all the buffers and reset all the finite state machines and counters on the board.
\\
The L0TP project has been developed using the VHDL hardware-description language in the Altera Quartus II integrated environment.

\section{Description of the L0TP modules}
In this chapter we describe the main modules of the L0TP: in the first paragraph, a description of the logic to receive the primitives from the NA62 detector is presented. Then, the details of the firmware to control the Etherent links are reported, followed by a description of the data-alignment performed by the L0TP. The algorithm to find a trigger matching condition, implemented using an associative memory, is then illustrated.
Finally the output logic to deliver the trigger information both to the  PC-Farm and to the NA62 detectors is described. 

\paragraph{Input primitives} \label{input}
Primitives are generated asynchronously by sub-detectors and sent to the L0TP during the entire beam burst period. 
A primitive is a data structure encoding the characteristics of an event in a detector, such as the energy in calorimeters or the hit-multiplicity and topology in the hodoscope. The time of the event completes the information. It consists of a timestamp, with a precision equal to the period of the master clock (24,95 ns) and a fine time, with the lowest significant bit corresponding to 1/256 of the main clock period (97.466 ps).
A variable number of L0 trigger primitives are merged into a single Multi-Trigger-Packet (MTP), characterized by a header word containing the data size and the identification code of the source generating the primitives.
\\
The time required to generate a primitive from a specific detector strongly depends on the complexity of the trigger algorithms implemented at the detector level.
This is the reason why the primitives related to the same physics event are asynchronous. 
The elapsed time for primitive generation is widely variable between sub-detectors: due to size limits for the L0TP memory buffers, this variable latency has been supported using additional constraints on trigger primitives:

\begin{itemize}
\item  primitives have to be packed to contain events in a certain time window (\textit{frame});
\item  frames have to be issued continuously, even if they are empty.
\end{itemize}

In NA62 the frame period is 6.4 $\mu$s, and MTP packets are sent periodically with the same period, using the UDP-Ethernet protocol.
Even with the frame structure, there are two kinds of delays that L0TP has to work with:
depending on the detector processing time, a fixed offset could be present between two different systems. 
As a result, the primitive related to a certain time could be stored in the packet $N$ for a faster detector, while the primitive related to the same event could be in the packet $M$ for a slower one, where $N<M$.
This delay remains fixed during the run and can be compensated by the L0TP logic, knowing how many frames are between the same event in the two subsystems.
\\
Moreover, the L0TP must guarantee enough margins to absorb any intrinsic fluctuation in the primitive production-time performed by the different sub-systems. Fluctuations depend on the complexity of the event, i.e. on the number of channels hit. 
The maximum timing fluctuation that can be accommodated is determined by the dimension of the L0TP buffers.

\paragraph{Ethernet module}
In order to receive/send information through the DE4 Ethernet links, an Internet Protocol/User-Datagram Protocol (IP/UDP) engine, called \textit{ethlink} module, has been developed. It is the main interconnection block between the FPGA internal data-processing logic and the external Ethernet physical (PHY) interface devices.
The ethlink module has been designed with the goal of small footprint on FPGA resources. This hardware IP/UDP stack can transmit and receive frames with a maximum payload length of 1500 bytes.
The ethlink module architecture is optimized to sustain the theoretical maximum throughput of 1-Gbit Ethernet standard without frame loss\footnote{The ethlink module manages up to 1488095 frames/s with minimum frame payload of 46 byte and 81274 frames/s with maximum frame payload of 1500 bytes.}. 
The IP/UDP stack uses a \textit{Class C} IP address space: the network nodes are allocated
using 8-bit IP host addresses with a corresponding static MAC addresses. However, special IP host addresses are reserved for dynamic MAC address assignments. This solution requires a very small amount of memory and enables the connection of host nodes having an immutable MAC address but a configurable IP address.
The ethlink module is implementing up to 8 Ethernet physical interfaces (\textit{MAC modules}); 
a single MAC module is configurable with a maximum of 15-inputs and 15-outputs hardware ports. A physical Ethernet interface is shared between different hardware ports: the transmission ports compete the external PHY interface using a round-robin scheduler; the received frames are dispatched to the corresponding receiving hardware port. 
The MAC port read logic is designed to receive and consume data at full 1-Gbit speed (125 MBytes/s), sustaining the Ethernet receiving traffic without memory buffer overflows.
The MAC module is connected to the external physical device using a configurable
interface module working in parallel mode (RGMII) or serial mode (SGMII).
The RGMII interface is controlled using a custom code module, while for the SGMII interface a standard core library building block has been used\footnote{The SGMII interface uses Altera Triple-Speed Ethernet MegaCore intellectual property (IP) core configured as 1000BASE-X/SGMII PCS with PMA mode}\cite{TSE}. 
Both RGMII and SGMII external physical interface standards are supported and can be mixed. 
Each MAC module implements one bidirectional port for the remote control and one input/output port for the reception/transmission of the data frames.
The ethlink module has been moved to different development environment versions and tested with different FPGA devices with practically no modifications.

\begin{figure}[h] 
\centering
\includegraphics[width=0.8\textwidth]{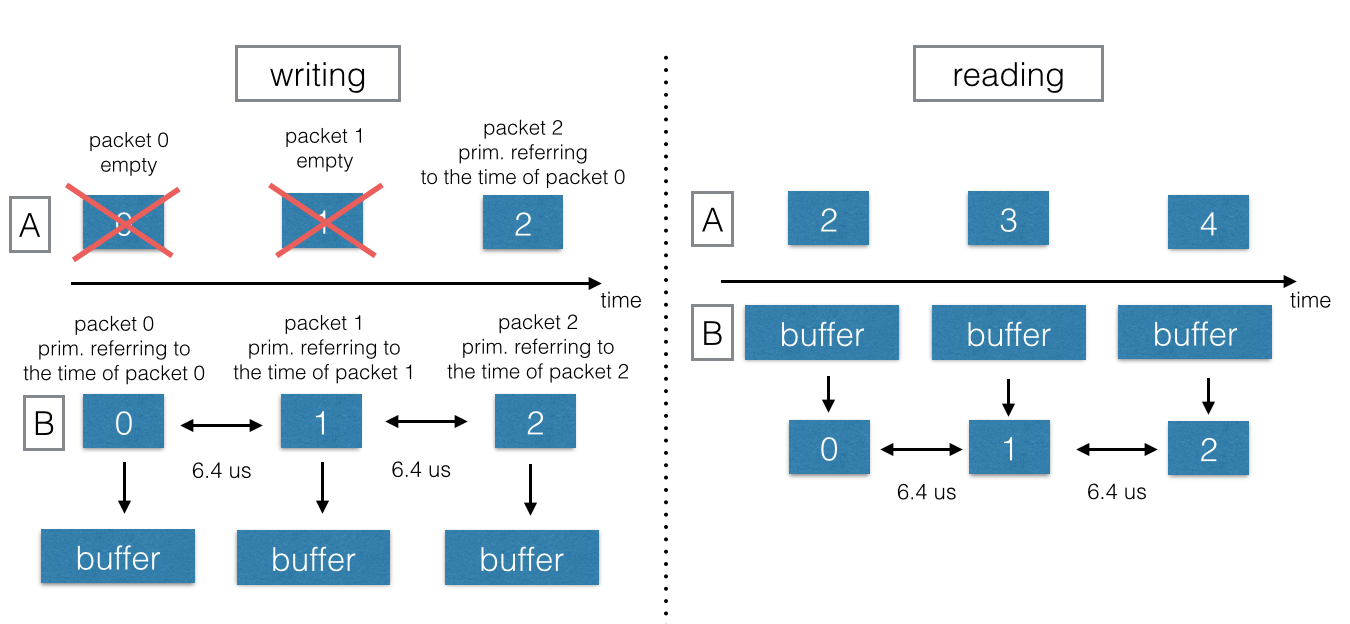}
\caption{The delay generator module is represented in this schematic. In this example, the detector A has a more complex algorithm to generate primitives respect to the detector B. Then, it spends more than 12.8 $\mu$s to produce the primitive referring to the time 0. For this reason a fixed offset of two packets is present in the streaming of data. The L0TP is programmed to store two packets of the B detector skipping them from A. When the third packet arrives, the reading process can start.}
\label{fig:picDelayGenerator}
\end{figure}

\paragraph{Alignment of the frames}
Starting from the header of the MTP, a finite state machine module extracts all the primitives coming from the ethlink module by processing the seven DE4-Ethernet links in parallel, thus handling almost 3 Gbit/s. 
The system also applies some time-consistency checks on the incoming primitives before accepting them, in order to comply with the frame structure.
The extracted information goes to a delay generator module to produce a programmable delay from the start of the burst, absorbing the potential fixed offset between the primitive sources, as described in the previous paragraph.
This module allows to realign the Ethernet-frame $N$ coming from detector \textit{i}, with the frame $M = N + \Delta N$ coming from detector \textit{j}. 
$\Delta N$ is the offset introduced by detector \textit{j} to process and generate the primitive related to the same event of the detector \textit{i}.
The L0TP stores in dedicated buffers the primitives from detector \textit{i} whilst skips $\Delta N$ frames from detector \textit{j}. 
In fact, buffering or skipping a single frame is equivalent to introduce a delay of 6.4 $\mu$s. 
When the first non-empty frame arrives from the slowest system, the delay generator module reads it from the Ethernet and, at the same time, reads the first frame stored in the buffer for the other detectors. When the next packets arrive, the L0TP reads again the slowest one from the Ethernet and the others from the buffers.
\\
The figure \ref{fig:picDelayGenerator} sketches how the delay generator module works.

The algorithm is based on a series of FIFOs, one for each detector. All the input primitives pass through these memories and, if the delay is set to zero, they are immediately read and they go to the next step of the logic. If the delay is greater than zero, only the FIFOs belonging to slower detectors are read and the output ignored, while data from the other sources are stored in the memory.
The FIFOs used to store primitives from faster sub-detectors are 8192-word deep, allowing to generate a compensation of 800 $\mu$s considering a mean primitive rate of 10 MHz. 
During the 2017 data taking  3-frame delay (19.2 $\mu$s) was introduced only for the calorimetric primitive generator, but larger values are expected to be required when including GPU-based primitive generators \cite{GPU}.

\paragraph{Alignment of the primitives}
When primitives are transferred from the delay module to the alignment logic, they are stored in RAMs, one for each sub-detector, using the time information to generate the address. 
The memory address is generated using the lowest significant bits of the timestamp and the most significant bits of the fine time. 
The number of  fine time bits used in the address is a parameter that can be set externally, matching a RAM location to a time interval (hereinafter called \textit{granularity}). Generating the address with the primitive time, the L0TP immediately performs a first rough alignment up to the granularity of the RAMs. At most, three fine time bits can be used, for a maximum granularity of 3.125 ns per memory slot, resulting in a maximum span for time alignment of 51 $\mu$s, which is equivalent to the maximum timing fluctuation that can be handled by the L0TP for each primitive generator.
During the 2017 data taking, the number of fine time used was 2, resulting in an accepted timing fluctuation of 102 $\mu$s. Nevertheless, each source was able to produce primitives with a maximum fluctuation of 25.6 $\mu$s. The choice of using a worst alignment was driven by the time resolution of the primitives coming from the sub-systems, close to 2.5 ns for the LKr-system.  
\\
In the L0TP implementation, two quantities were constraining the access to the RAM: the writing operation rate, which is determined by the primitive rate and the reading operation rate, which corresponds to the granularity.  In NA62, the write-access rate is such that the addresses of the RAM occupied by a primitive within a given frame are below 4\%. The remaining addresses are empty or filled with old data due to the circular-buffer organization of RAM.
To scan all the addresses of the RAM would have been much more demanding requiring: a read rate at least of 330 MHz, followed by several operations to decide if a primitive belonged to the present frame.
It was then introduced the concept of a reference detector. 
The memory addresses of the reference detector primitives are stored into a dedicated FIFO, marking the only memory locations which will be read for all sources. All the other sub-systems are checked to be in time with respect to the reference detector in a certain (programmable) time window, and then their conditions will be compared with the pre-selected trigger masks. 
To maintain complete flexibility, the choice of the reference detector is programmable by the user.
Being the reference detector always present by construction in all the trigger masks used during the data taking, it could be impossible to measure its efficiency. 
To solve this problem, a control, minimum-bias, module to generate triggers has been implemented.
This solution offers the possibility to create two different sample of events, one driven by the reference detector and the other by the control one. 
The two samples are uncorrelated streams of data in the FPGA, but they are correlated from the point of view of physics, so that using the control sample it is possible to measure the efficiency of each trigger mask. 
The control system uses primitives coming from the NA48 old-hodoscope, still present in the NA62 detector system.

\paragraph{Reading process}
Depending on how many bits of the fine time are used, the read pointer of the RAM goes to the location hosting the primitives of the reference or control detector. 
All the RAM location of the seven sources are read in parallel.
The implemented alignment logic uses RAMs as circular buffers with finite size. During the reading process, The L0TP simply analyzes primitive data without clear the RAMs content. To avoid reprocessing old data, the MSBs of the timestamps are written in the memory slots. This additional information gives the possibility to discriminate between old and new primitives: only when the MSBs coincide the primitives are then kept, otherwise they are discarded.
\\
To avoid reprocessing previous primitives, the MSBs of the timestamps are written in the memories. 
During the reading process they are compared. When the MSBs coincide, the primitives are then kept, otherwise they are discarded.
\\
Primitives of the same physics event may have different timing and thus occasionally fall in adjacent time slots (also referred to as "edge effect").
To avoid it, the previous and the next slot around the reference location are also read.
When primitives are read from the RAMs, a tighter timing cut is applied around the reference detector time, using the entire fine time information.
Only if the time difference between the reference detector is lower than a pre-selected value the primitive goes to the Associative Memory Module. 
Different timing-cut can be applied to the different sources, depending on their online timing resolution.
In principle, coincidences to 100 ps are allowed. Due to the online time-resolution of the NA62 detectors, the timing-cut applied are of the order of 5 ns with respect to the reference detector time. 

\paragraph{Associative Memory Module}
The Associative Memory Module (AMM) is implemented in two parts: the first part is a shift register to store the information of the three RAM slots read from each detector.
A logical OR between the Primitive IDs of the three slots is performed, generating a global primitive ID with all the in-time information.
The result of the OR is then moved to the second part which works like an associative PROM, comparing the global primitive IDs of all the sources against a table of stored masks.
The masks are constructed following the user requests through a L0TP interface and it is basically the combination of the primitive IDs that should be satisfied by a trigger. 
Each bit of a trigger mask can be set as: requested, not requested ($e.g.$ for a veto conditions), ignored.
All the masks are processed in parallel in one clock cycle. 
The L0TP allows to have a maximum of 16 independent masks.
\\
In order to fulfill the NA62 output bandwidth, each mask can be down scaled of a factor selected by the user.
The output bus of the AMM brings all the information useful to an offline reconstruction of the primitives that have generated the trigger.

\paragraph{Output triggers}
When data satisfy the selection of the AMM masks, the trigger signal is delivered to all the sub-detectors after programmable fixed latency. It can be set up to 1 ms.
To generate the latency, a 1 ms deep circular buffer is implemented. The trigger information is stored in the buffer using the least significant 16 bits of the timestamp, preventing to loose the time sorting. 	
A counter starts when the start of burst signal arrives, and it counts with the frequency of 40 MHz, while the read address pointer of the circular buffer remains idle. 
When the latency time is reached, the reading pointer moves address by address in a row, allowing to read the position \textit{X} at the time \textit{X+latency}.
When a not-empty slot is found, the information is send to all detectors, and also to the NA62 PC-Farm, to reconstruct offline all the characteristics of the triggering event.
\\
A minimum dead time of 75 ns between two consecutive triggers is required due to the trigger transmission logic of NA62; in fact the trigger distribution performed by the TTCex board requires 3 clock cycles to receive, register and deliver a trigger signal to the detectors.   
For this reason the L0TP memorizes the last trigger delivered and checks the time difference between two consecutive output-signals.  
\\
Together with the primitives, the L0TP handles asynchronous choke/error signals from sub-detectors, to monitor if any overload condition has been reached. During the period in which the choke/error is active, no L0 triggers are dispatched. In order to guarantee the data acquisition efficiency and integrity at all times, both the start and the end of the choke/error condition are signaled to all systems by delivering a special trigger, which has to be acknowledged. 
Random, periodics and calibration triggers are also available to test and monitor the various NA62 systems (see special trigger paragraph).
The total rate in the data taking cannot be higher than 1 MHz. 
In case beam intensity fluctuations resulted in instantaneous rates above 1 MHz, an \textit{autochoke} protection mechanism has been implemented, forcing the L0TP to stop dispatching triggers and and to communicate the status to all the detector.
 
\paragraph{Special triggers}
In the NA62 implementation of the L0TP project, different special triggers can be used to monitor the status of the data taking, to test the acquisition system and to calibrate the detectors.
The first special trigger generator is the periodic one. It allows to have two different periodic trigger flows with two independent periods. The period has to be a multiple of the global clock period, roughly 25 ns in NA62.
It is possible to set the starting and the ending point with respect to the start of the burst. 
NA62 exploits the periodic triggers to detect noisy pixels in the GigaTracker spectrometer and to monitor the pedestal values of the LKr electromagnetic calorimeter. 
\\
The random generator is implemented using linear feedback shift registers. 
When a random number has been generated, if the LSB is equal to 1, a trigger signal is generated, otherwise another number is produced.
The frequency of the number generation can be set by the user, determining the mean frequency of the random triggers.
As the periodic generator, it has the possibility to set the starting point during the burst.
In NA62, the random trigger generator is used to test the data acquisition performances of the experiment.
\\
Finally, several detectors need special triggers for calibration purposes.
The L0TP can handle these special triggers: when a given detector sends primitives flagged with the most significant bit (bit 15) equal to 1, the L0TP logic accepts them independently on the configuration of the other systems.
For these events, the primitive information is written in a FIFO as soon as they are extracted from the delay generator.
When the FIFO is not empty, a trigger is generated and sent directly to the output stage, skipping all the alignment and matching logic.
\\
In order to deal with LKr calibration runs, the L0TP Board was adapted to handle NIM signals indicating a calibration event in the calorimeter. 
The L0TP creates an output string latching the internal timestamp to the trigger.
This event is then propagated to the output stage and written in the memories in order to be dispatched to detectors and PC-farm after the fixed latency. 

\begin{figure}
\centering
\includegraphics[width=0.6\textwidth]{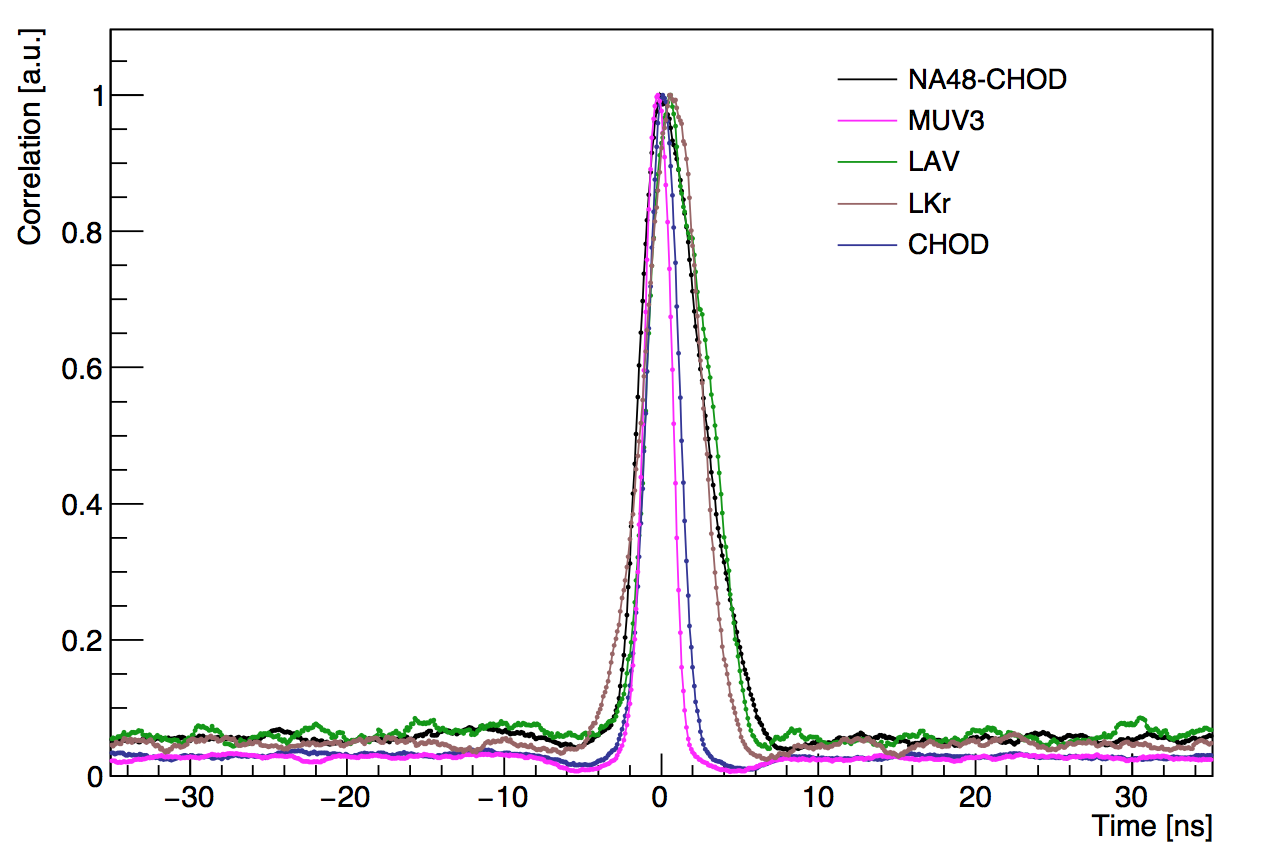}
\caption{The figure shows the timing-alignment between different sub-detectors prior any correction. The width of each distribution represents the time resolution of a given detector as measured online.}
\label{fig:Correlation}
\end{figure}

\paragraph{Input checks}
The use of the UDP standard packets to deliver primitives grants the implementation of parallel tools to check of the inputs to the trigger: seven switches\footnote{D-Link Gigabit Ethernet EasySmart Switch DGS-1100} with a mirrored port have been installed between the source and the DE4 board to duplicate the trigger-primitive stream going from the detectors to the L0TP. The second data stream is sent to a stand-alone workstation for a complete analysis.
The analysis includes a continuous monitoring of the beam quality  (\textit{i.e.} intensity, length, modulations) and of the detector outputs (\textit{i.e.} rates, Primitive ID compositions, timing-alignment).
In figure \ref{fig:Correlation} the timing-alignment between different sub-detectors is shown as monitored by the stand-alone workstation, before timing corrections. The timing is measured with respect to the RICH, set as reference detector.
\section{Performances}
\begin{figure}
\centering
\includegraphics[width=0.6\textwidth]{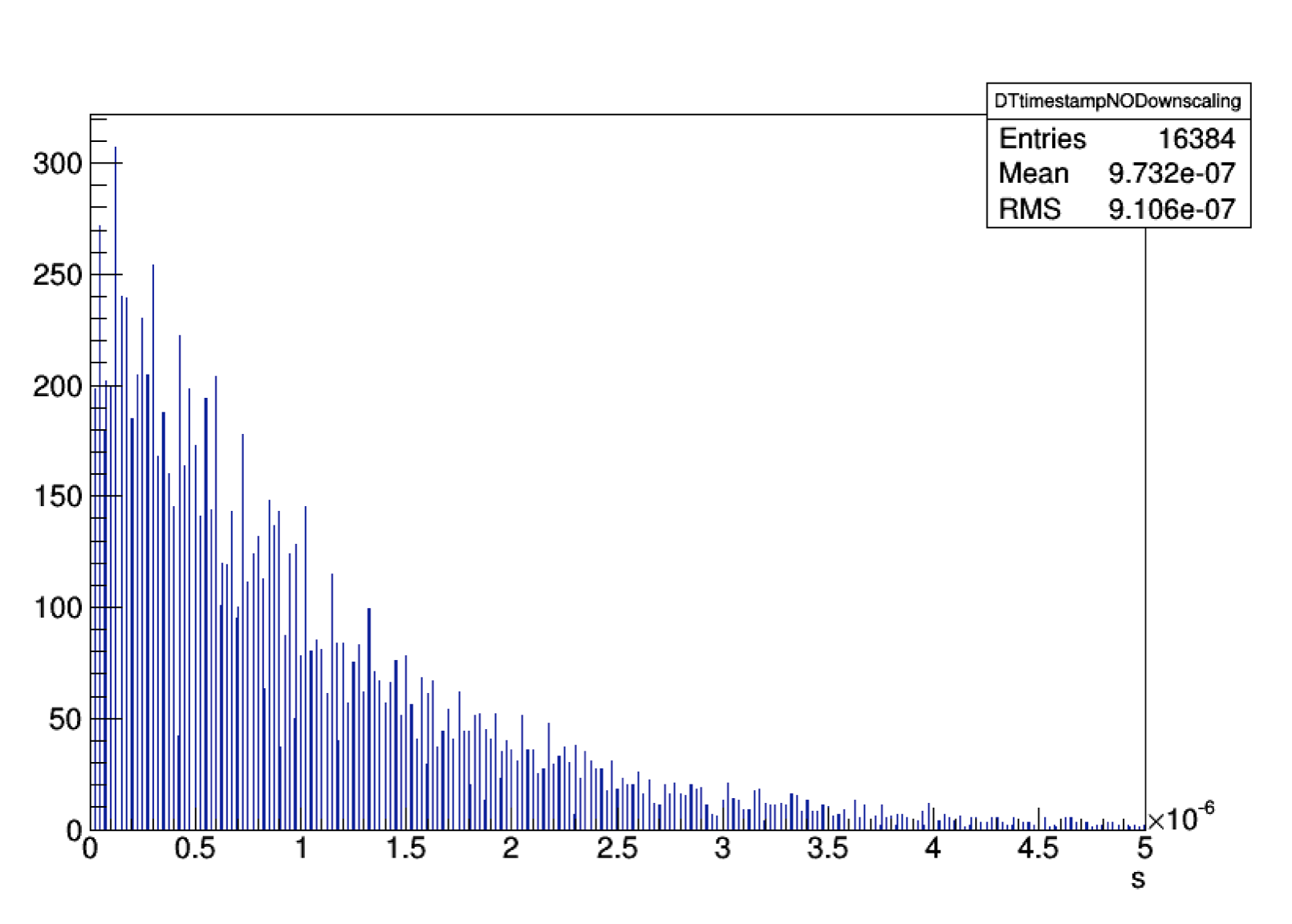}
\caption[Random Primitive]{The distribution shows the time difference between consecutive primitives generated with an exponential-random distance between timestamps. Loading the primitives in the Simulator memory was possible to simulate a realistic condition of data acquisition.}\label{fig:RandomPrimitive}
\end{figure}

\begin{figure}[h!]
\centering
  \includegraphics[width=0.6\textwidth]{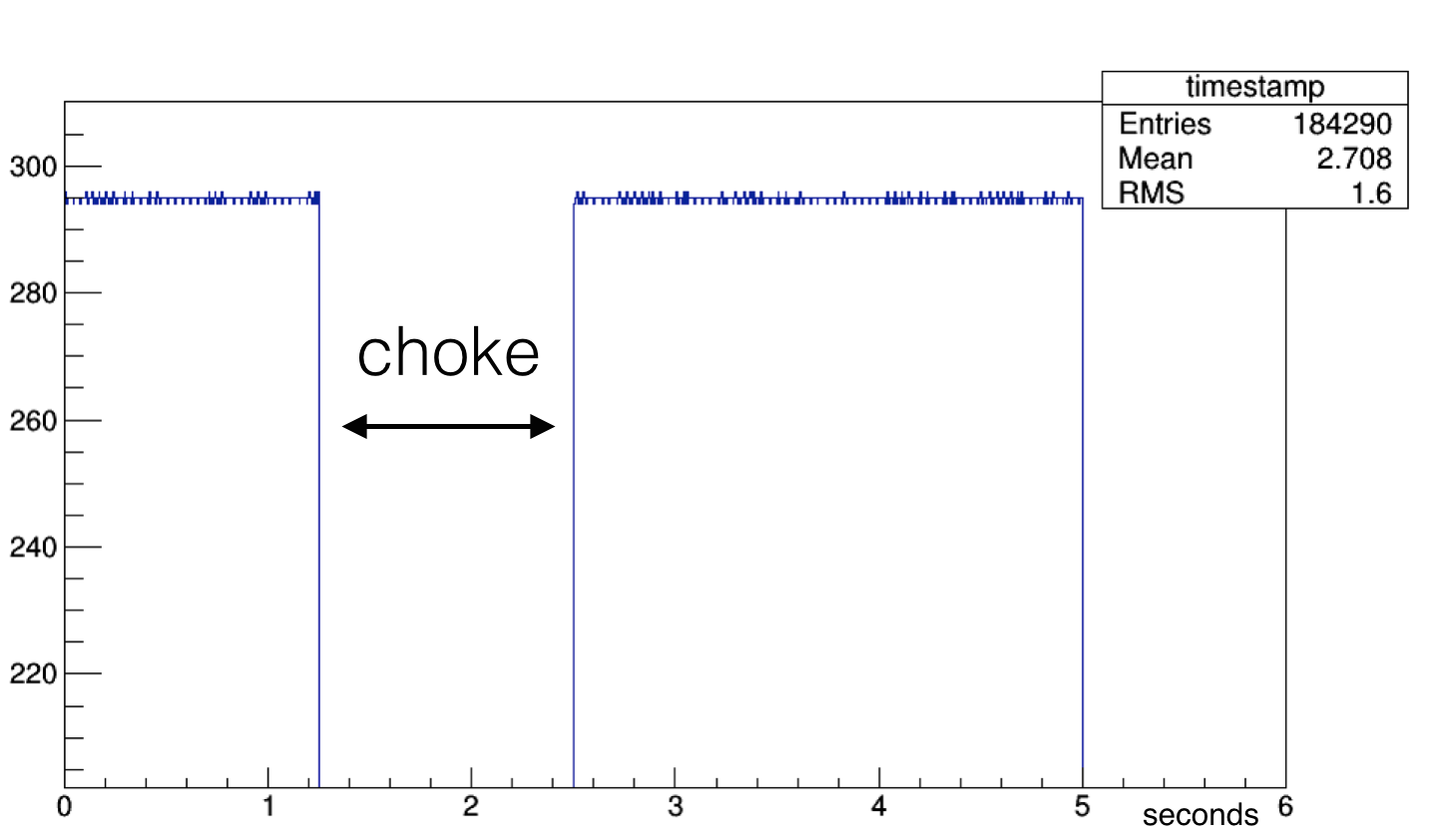}
  \caption{The distribution depicts the trigger arrival time during burst. Whenever a choke (error) condition has been generated, the L0TP inhibits the output data stream towards the detectors and the PC-Farm. The trigger stream restarts as soon as the choke (error) condition is over.}
  \label{fig:ChokeFig}
\end{figure}

Since the beginning of the project, the L0TP logic has been extensively tested in a complete test-bench in order to verify that coincidences between different sources are indeed found and that the data transmission to both farm-PCs and detectors is performed as expected.
The test-bench has been implemented by connecting two different DE4 boards, one hosting the L0TP firmware, while the second board, called \textit{Simulator}, emulating the experimental environment surrounding the L0TP. In particular it hosted:
\begin{itemize}
\item a 40 MHz external clock generator;
\item a choke/error signal generator;
\item a start/end of burst signal generator;
\item a primitive dispatcher module.
\end{itemize}
The Simulator was receiving the triggers back from the L0TP.
To check the correctness of the triggers received, they were time-stamped with the 40 MHz clock and results were retrieved through Ethernet connections to a workstation where the data were monitored and compared with the expected values.
\\
At the initial stage of the project, when the parallel acquisition system was not available, the simulated primitives have been used to test the L0TP logic.
Given the value of the (N-1)th timestamp, the next timestamp, Nth, has been generated with an exponential distribution as in equation \ref{eqtimestamp}.

\begin{eqnarray}
Timestamp_N &=& Timestamp_{N-1} \nonumber \\
&+& (-frequency \times log(1-(finetime))
\label{eqtimestamp}
\end{eqnarray}

where \textit{frequency} is the average frequency for the primitive generation and the finetime has been generated with a random flat distribution. The distribution of the time difference between consecutive primitives is shown in figure \ref{fig:RandomPrimitive}.
The behavior of the system to handle the choke/error signals is shown in figure  \ref{fig:ChokeFig}. 
Performing this test, it was verified that while a choke (or error) signal was received, the L0TP dispatched a special trigger to indicate the timestamp in which the choke signal occurred.
Then the logic to send triggers to the PC-Farm was inhibited and immediately restarted after the end of the choke/error signal. 
Special triggers as choke and errors are sent without delay in order to act immediately in reducing the pressure of data requests on detector side.
\begin{figure}[h!]
\centering
 \includegraphics[width=0.6\textwidth]{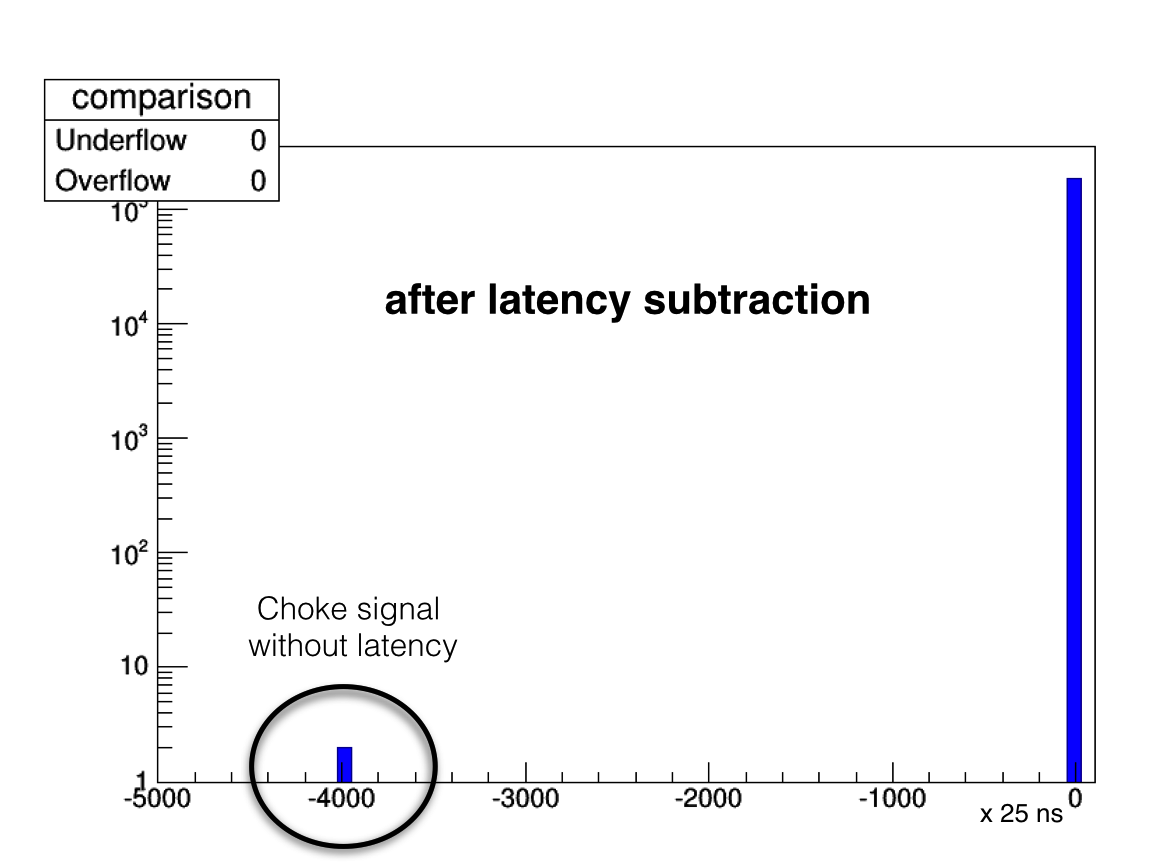}
 \caption{Distribution of the time difference between timestamps of data sent to PC-Farm and detectors, after latency subtraction. Choke (error) signals are delivered without any latency. For this reason two events (\textit{choke on} and \textit{choke off} signals) are located in bin -4000, corresponding to 100 $\mu$s, the preset latency value.}
\label{fig:ChokeTriggerwordFig}
\end{figure}
\\
Figure \ref{fig:ChokeTriggerwordFig} shows the difference of the timestamp between triggers sent to the PC-farm  and triggers sent to the detectors after the latency subtraction: all triggers are correctly handled by the L0TP and there are no differences between the times sent to PCs and detectors. Only two events (\textit{choke on} and \textit{choke off} signals) are located in bin -4000, which corresponds to 100 $\mu$s, correctly delivered without latency.
\\
After the implementation of the primitive parallel acquisition system in NA62, it has been possible to record the primitives generated by detectors during the data taking and replicate part of those bursts on a test stand, reproducing the same characteristics of the trigger dispatchers.   
\begin{figure}[h!]
\centering
  \includegraphics[width=0.6\textwidth]{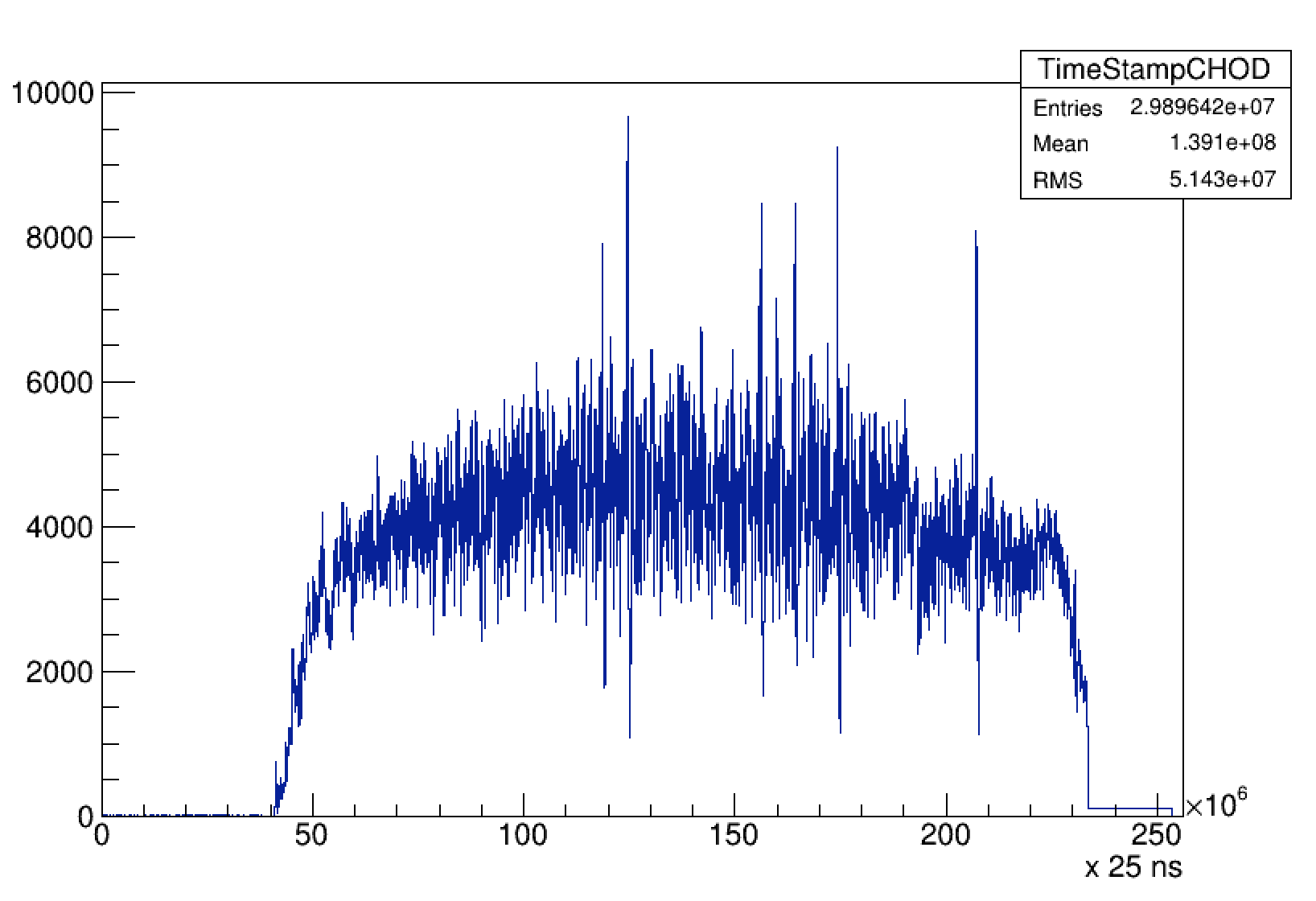}
  \caption{Distribution of the NA48-CHOD primitive timestamp used for tests. The primitives have been loaded in the Simulator memory and sent to the L0TP board, reproducing the real burst conditions.}
  \label{fig:chodmuvtimestamp}
\end{figure}
Figure \ref{fig:chodmuvtimestamp} shows the timestamp profiles of the primitives generated by NA48-CHOD and replicated by the test bench. Together with NA48-CHOD, the Simulator replicates also the primitives of MUV3 and LAV, allowing to test the algorithm to find the coincidences between the different sources.
\\
\begin{figure}
\centering
\includegraphics[width=0.5\textwidth]{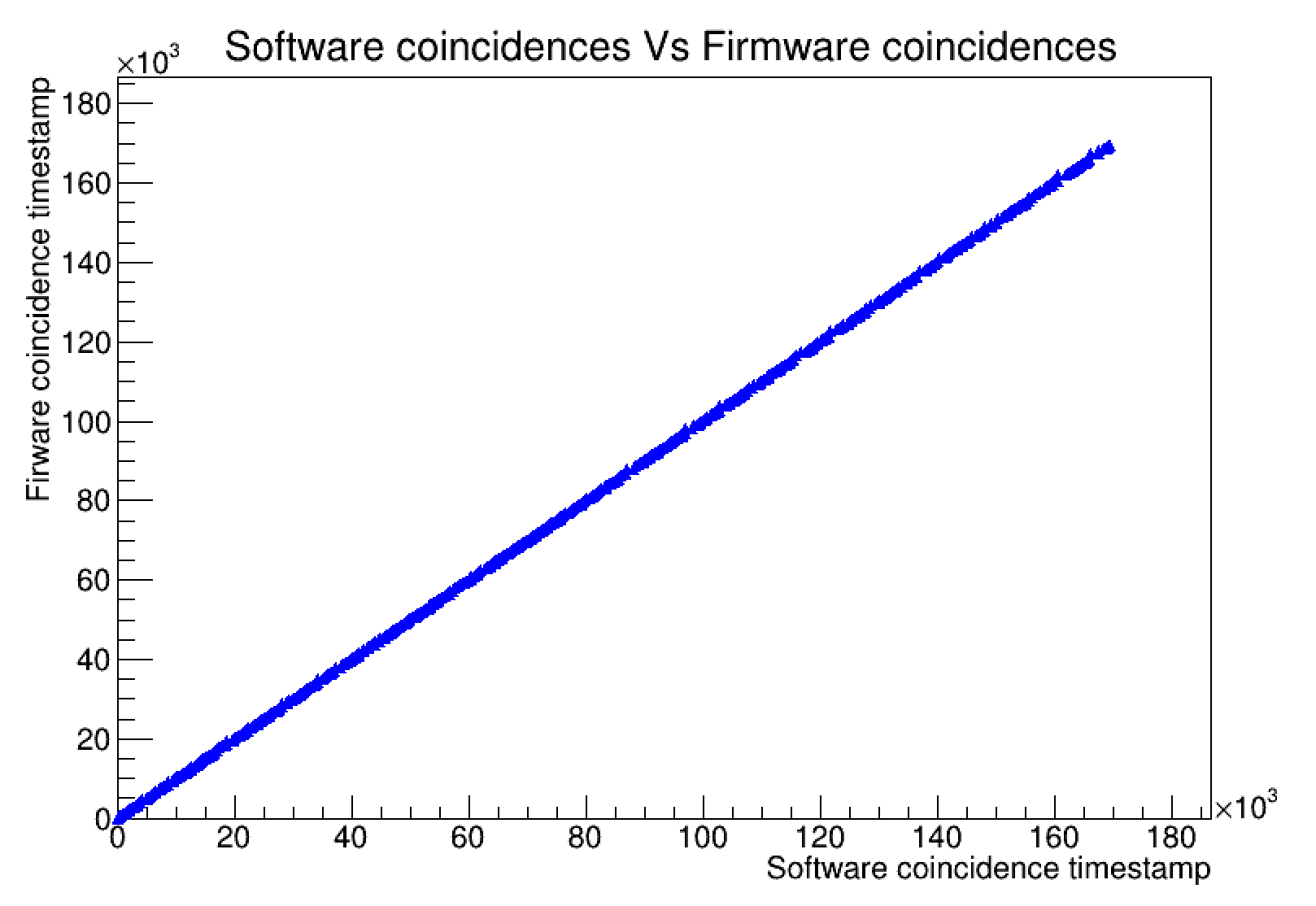}
\caption{Distribution of the timestamps found by the L0TP firmware versus the timestamps found by the software simulation of the L0TP algorithm. All the events have been found by the L0TP.}
\label{fig:SoftwareVsFirmware}
\end{figure}
The sample of triggers identified by the L0TP algorithm and the one obtained using a software-emulator program (written in $C++$ language) of the trigger logic are fully overlapping, as shown in figure \ref{fig:SoftwareVsFirmware}. 
From the limited statistics that have been analyzed one can evaluate an upper limit to the inefficiency of the L0TP logic, resulting less than $1.2 \times 10^{-3}$. 
\\
The L0TP is included in the NA62 data acquisition system since 2014, and in 2017 the state of the art of the system has been commissioned, sustaining an average input rate up to 10 MHz from each detector without experiencing any issue.

\section{Conclusions}
The work presented in this paper focuses on the L0 Trigger Processor of the NA62 experiment.
The algorithm implemented on the FPGA and the environment set up for the proper trigger operation have been presented and the capability to achieve an efficient data taking in the challenging environment of very high particle flux of the NA62 beam have been reported. 
The L0TP reached the design input/output rates guaranteeing a smooth data taking thanks to the stability of system.

\acknowledgments
The authors are grateful to the whole NA62 collaboration for fruitful discussions and suggestions during commissioning and early operation of the L0TP. Special thanks are due to the Ferrara and Turin teams for their trusting and long-lasting support. This work was the subject of the PhD thesis of one of the authors (D.S.).


\end{document}